%% file: PQ_DPF2019.tex
\documentclass[11pt]{article}
\usepackage[margin=1in]{geometry}
\usepackage{lipsum}
\usepackage{epsfig}
\usepackage{amsmath}
\usepackage{amssymb}
\usepackage{amsfonts}
\usepackage{multicol}
\usepackage{graphicx,url}
\usepackage[english]{babel}   
\usepackage[utf8]{inputenc}
\usepackage{xfrac}
\usepackage{multirow}
\usepackage{geometry}
\usepackage{hyperref}
\usepackage{bbm}

\input econfmacros.tex 

\def\Title#1{\begin{center} {\Large {\bf #1} } \end{center}}
\def\Author#1{\begin{center} {\normalsize {\sc #1} } \end{center}}
\def\Institution#1{\begin{center} {\normalsize {\it #1} } \end{center}}
\def\Abstract#1{\noindent {\normalsize {\bf Abstract:} {\normalfont #1}}}
\def\Conference{\vspace{4mm}\begin{raggedright} {\normalsize {\it Talk presented at the 2019 Meeting of the Division of Particles and Fields of the American Physical Society (DPF2019), July 29--August 2, 2019, Northeastern University, Boston, C1907293.} } \end{raggedright}\vspace{4mm}}

\begin{document}

%
%

\Title{A $U(1)_{X}$ extension to the SM with three families and Peccei Quinn symmetry}

\Author{Y.A. Garnica, R. Martinez}

\Institution{Departamento de F\'{i}sica\\ Universidad Nacional de Colombia\\
 Ciudad Universitaria$,$ K. 45 No. 26-85$,$ Bogot\'a D.C.$,$ Colombia}

\Abstract{We propose a non-universal $U(1)_{X}$ extension to the Standard Model with three families and an additional global anomala Peccei-Quinn (PQ) symmetry. The breaking of the former allows us to give masses to the exotic fermionic sector and the later generates the necessary zeros in the mass matrices to explain the fermionic mass hierarchy. In addition, the large energy scale associated with the spontaneously breaking (SSB) of the PQ symmetry provides a solution to the strong CP-problem and an axion that could be a possible dark matter candidate. Also, the SSB allows to generate right-handed neutrino masses, so the active neutrinos acquire $eV$-mass values due to the see-saw mechanism implementation. }

\Conference

%
%

\section{Introduction}

A $U(1)_{X}$ extension of the Standard Model (SM) with a $U(1)_{PQ}$ symmetry is built and explored by considering a non-universal $X$ charge assignment that makes the model anomaly free with the inclusion of additional quarks and leptons. In \cite{Mantilla} a $Z_{2}$ symmetry was introduced to produce an ansatz for the fermion mass matrices. In the present version we have replaced the discrete symmetry by a PQ symmetry. This election produces the same Yukawa lagrangian, also requiring the use of two Higgs doublets and generating solution to the strong CP-problem. This two Higgs doublets are considered to break the electroweak symmetry and give masses to the top and bottom quarks at tree level. The singlet $\chi$ with vacuum expectation value (VEV) $v_{\chi}$ produces the breaking of the horizontal symmetry and generates masses of the exotic particles and singlets. We also consider two additional scalar fields; $\sigma$ is necessary to produce the masses of the lightest fermions through radiative corrections and the scalar $S$ causes the breaking of the PQ symmetry and gives masses to the right-handed neutrinos allowing to generate masses through a see-saw mechanism to the lighter active neutrinos.

\section{Scalar sector}
\begin{table}[h]
\begin{center}
\begin{tabular}{ccc}
Scalar bosons	&	$X$	&	$U(1)_{PQ}$	\\ \hline 
\multicolumn{2}{c}{Higgs Doublets}\\ \hline\hline
$\phi_{1}=\left(\begin{array}{c}
\phi_{1}^{+} \\ \frac{h_{1}+v_{1}+i\eta_{1}}{\sqrt{2}}
\end{array}\right)$	&	$2/3$	&	$x_1$	\\
$\phi_{2}=\left(\begin{array}{c}
 \phi_{2}^{+} \\ \frac{h_{2}+v_{2}+i\eta_{2}}{\sqrt{2}}
\end{array}\right)$	&	$1/3$	&	$x_2$	\\   \hline\hline
\multicolumn{2}{c}{Higgs Singlets}\\ \hline\hline
$\chi  =\frac{\xi_{\chi}  +v_{\chi}  +i\zeta_{\chi}}{\sqrt{2}}$	& $-1/3$	&	$x_{\chi}$	\\   
$\sigma$	& $-1/3$	&	$x_\sigma$	\\  
$S =\frac{\xi_{S}+v_{S}+i\zeta_{S}}{\sqrt{2}}$	& $-2/3$	&	$x_S$	\\
\hline \hline 
\end{tabular}
\caption{Non-universal $X$ quantum number and $U(1)_{PQ}$ for Higgs fields.}
\label{tab:Scalar-content}
\end{center}
\end{table}
In the present model, there are two inherent large scales: the first is associated with the SSB of the PQ symmetry, which provides a solution to the strong CP-problem and generates a viable candidate for dark matter (axion). The other is related to the $\chi$ singlet with VEV $v_{\chi}$ which allows the spontaneous breaking of the $U(1)_{X}$ symmetry as mentioned above. The hierarchy between the scales is $v_{S}\gg v_{\chi}\gg v$, where $v$ corresponds to the electroweak VEV $v=\sqrt{v_{1}^{2}+v_{2}^{2}}\sim 246 GeV$, $v_{\chi}\sim TeV$ and $v_{S}\sim 10^{10}GeV$ is the VEV of the $S$-singlet. The implementation of an anomalous PQ symmetry through a DFSZ model \cite{DFS1981} generates the ansatz for the mass matrices of fermions, giving an explanation for hierarchy values and mixing angles. The assignment of the PQ charges generates several restriction on the scalar potential, that can be expressed as:

\begin{align}
V &= \mu_{1}^{2}\phi_{1}^{\dagger}\phi_{1} + \mu_{2}^{2}\phi_{2}^{\dagger}\phi_{2} + \mu_{\chi}^{2}\chi^{*}\chi + \mu_{\sigma}^{2}\sigma^{*}\sigma+\mu_{S}S^{*}S
 + \lambda_{1}\left(\phi_{1}^{\dagger}\phi_{1}\right)^{2} \nonumber\\&
 + \lambda_{2}\left(\phi_{2}^{\dagger}\phi_{2}\right)^{2} 	
 + \lambda_{3}\left(\chi^{*}\chi \right)^{2} + \lambda_{4}\left(\sigma^{*}\sigma \right)^{2} 	
 + \lambda_{5}\left(\phi_{1}^{\dagger}\phi_{1}\right) \left(\phi_{2}^{\dagger}\phi_{2}\right) \nonumber\\&
 + \lambda'_{5}\left(\phi_{1}^{\dagger}\phi_{2}\right)\left(\phi_{2}^{\dagger}\phi_{1}\right)	
 + \left(\phi_{1}^{\dagger}\phi_{1}\right)\left[ \lambda_{6}\left(\chi^{*}\chi \right) + \lambda'_{6}\left(\sigma^{*}\sigma \right) 
 \right] 	\nonumber\\&
 + \left(\phi_{2}^{\dagger}\phi_{2}\right)\left[ \lambda_{7}\left(\chi^{*}\chi \right) + \lambda'_{7}\left(\sigma^{*}\sigma \right) \right] 	
 + \lambda_{8}\left(\chi^{*}\chi \right)\left(\sigma^{*}\sigma \right) +\lambda_{9}(S^{*}S)^2     \nonumber\\&
 + (S^{*}S)\left[\lambda_{10}\left(\phi_{1}^{\dagger}\phi_{1}\right)+\lambda_{11}\left(\phi_{2}^{\dagger}\phi_{2}\right)+\lambda_{12}\left(\chi^{*}\chi\right)+\lambda_{13}\left(\sigma^{*}\sigma\right)\right]  \nonumber\\&                                +\lambda_{14}\left(\chi S^{*}\phi_{1}^{\dagger}\phi_{2}+h.c.\right),
\label{eq:scalar-potential}
\end{align}
where the term proportional to $\lambda_{14}$ is necessary to avoid trivial PQ charges for the scalar sector and taking $\mu_{\sigma}^{2}>0$ to avoid VEV of the $\sigma$ fields.

After symmetry breaking all Higgs doublets and singlets (except for $\sigma$) acquire a VEV which allows us to construct a squared mass matrix $M_{R}^{2}$ for CP-even scalar particles in the  $(h_{1},h_{2},\xi_{\chi},\xi_{S})$ basis: 
\begin{equation}
    M_{R}^{2}=
    \begin{pmatrix}
    \lambda_{1}v_{1}^{2}-\dfrac{\lambda_{14}}{4}\dfrac{v_{2}v_{\chi}v_{S}}{v_{1}}&\dfrac{\bar{\lambda}_{5}}{2}v_{1}v_{2}+\dfrac{\lambda_{14}}{4}v_{\chi}v_{S}&\dfrac{\lambda_{6}}{2}v_{1}v_{\chi}+\dfrac{\lambda_{14}}{4}v_{2}v_{S}&\dfrac{\lambda_{14}}{4}v_{2}v_{\chi}+\dfrac{\lambda_{10}}{2}v_{1}v_{S}\\
    *&\lambda_{2} v_{2}^{2}-\dfrac{\lambda_{14}}{4}\dfrac{v_{1}v_{\chi}v_{S}}{v_{2}}&\dfrac{\lambda_{7}}{2}v_{2}v_{\chi}+\dfrac{\lambda_{14}}{4}v_{1}v_{S}&\dfrac{\lambda_{14}}{4}v_{1}v_{\chi}+\dfrac{\lambda_{11}}{2}v_{2}v_{S}\\
    *&*&\lambda_{3}v_{\chi}^{2}-\dfrac{\lambda_{14}}{4}\dfrac{v_{1}v_{2}v_{S}}{v_{\chi}}&\dfrac{\lambda_{14}}{4}v_{1}v_{2}+\dfrac{\lambda_{12}}{2}v_{\chi}v_{S}\\
    *&*&*&\lambda_{9}v_{S}^{2}-\dfrac{\lambda_{14}}{4}\dfrac{v_{1}v_{2}v_{\chi}}{v_{S}}
    \end{pmatrix}.
\end{equation}
This matrix has Rank $M_{R}^{2}=4$ and we define $\bar{\lambda}_{5}=\lambda_{4}+\lambda_{5}$. In order to obtain the eigenvalues, we use the VEV hierarchy $v_{S}\gg v_{\chi}\gg v$ to calculate them perturbatively. Through the scaling of couplings in the scalar potential in eq.  (\ref{eq:scalar-potential}), it is possible to made our model technically natural generating an explicit decoupling between SM and the neutral singlet on the PQ-scale. So, requiring the relations:
\begin{align*}
    \lambda_{6}&\equiv a_{6}\frac{v_{1}^{2}}{v_{\chi}^{2}},\quad \lambda_{7}\equiv a_{7}\frac{v_{2}^{2}}{v_{\chi}^{2}},\quad \lambda_{10}\equiv a_{10}\frac{v_{1}^{2}}{v_{S}^{2}},\quad \lambda_{11}\equiv a_{11}\frac{v_{2}^{2}}{v_{S}^{2}}\\
    \lambda_{12}&\equiv a_{12}\frac{v_{\chi}^{2}}{v_{S}^{2}},\quad \lambda_{14}\equiv a_{14}\frac{v^{2}}{v_{\chi}v_{S}},
\end{align*}
is possible to build a natural hierarchy between the PQ and electroweak scale, without unpleasant fine tuning \cite{Bertolini}.
The leading-order contribution to the $M_{R}^{2}$-matrix is giving by:
\begin{equation}
\mathit{M}_{\mathrm{R}}^{2}\approx
\begin{pmatrix}
\lambda_{1}v_{1}^{2}-\dfrac{c_{14}}{4}\dfrac{v^{2}v_{2}}{v_{1}}&\dfrac{\bar{\lambda_{5}}}{2}v_{1}v_{2}+\dfrac{c_{14}}{4}v^{2}&0&0\\\dfrac{\bar{\lambda_{5}}}{2}v_{1}v_{2}+\dfrac{c_{14}}{4}v^{2}&\lambda_{2}v_{2}^{2}-\dfrac{c_{14}}{4}\dfrac{v^{2}v_{1}}{v_{2}}&0&0\\
0&0&\lambda_{3}v_{\chi}^{2}&0\\0&0&0&\lambda_{9}v_{S}^{2}
\end{pmatrix}.
\end{equation}
 So, at LO, the heaviest eigenvalues are decoupled from the electroweak scale and keeping only $\mathcal{O}(v^{2})$ terms, the eigenvalues of the CP-even sector can be listed as:
 \begin{equation}
     \left\{\mathcal{O}(v^{2}),\mathcal{O}(v^{2}),\lambda_{3}v_{\chi}^{2},\lambda_{9}v_{S}^{2}\right\},
 \end{equation}
 where the lightest corresponds to the Higgs boson with mass of $125 GeV$.

The pseudo-scalar matrix in the basis  $(\eta_{1},\eta_{2},\zeta_{\chi},\zeta_{S})$ is giving by:

\begin{equation}
    M_{I}^{2}=-\dfrac{\lambda_{14}}{4}
    \begin{pmatrix}
    \dfrac{v_{2}v_{\chi}v_{S}}{v_{1}}&-v_{\chi}v_{S}&-v_{2}v_{S}&v_{2}v_{\chi}\\
    -v_{\chi}v_{S}&\dfrac{v_{1}v_{\chi}v_{S}}{v_{2}}&v_{1}v_{S}&v_{1}v_{\chi}\\
    -v_{2}v_{S}&v_{1}v_{S}&\dfrac{v_{1}v_{2}v_{S}}{v_{\chi}}&-v_{1}v_{2}\\
    v_{2}v_{\chi}&-v_{1}v_{\chi}&-v_{1}v_{2}&\dfrac{v_{1}v_{2}v_{\chi}}{v_{S}}
     \end{pmatrix}.
     \label{Cp-odd_sector}
\end{equation}

 This matrix contains three zero-mass modes, two corresponding to the would-be Goldstone bosons eaten by the $Z,Z'$ and one corresponding to the axion that acquires mass by non-perturbative QCD. The massive state corresponding to the $A_0$ pseudoscalar boson is:
\begin{equation}
    m_{A^0}^{2}=\dfrac{\lambda_{14}}{4}\left(\frac{v_{\chi}v_{S}}{2s_{2\beta}}+\frac{v^{2}(v_{\chi}^{2}+v_{S}^{2})s_{2\beta}}{2v_{\chi}v_{S}}\right).
\end{equation}
For the charged scalar sector, we have the rank 1 matrix:
\begin{equation}
    M_{C}^{2}=\dfrac{1}{4}
    \begin{pmatrix}
    \lambda_{5}v_{2}^{2}-\lambda_{14}\dfrac{v_{2}v_{\chi}v_{S}}{v_{1}}&\lambda_{5}v_{1}v_{2}+\lambda_{14}v_{\chi}v_{S}\\
    *&\lambda_{5}v_{1}^{2}-\lambda_{14}\dfrac{v_{1}v_{\chi}v_{S}}{v_{2}}    \end{pmatrix},
\end{equation}
which implies one would-be Goldstone associated to the $W_{\mu}^{\pm}$ boson and one charged Higgs with mass equal to: 
\begin{equation}
    m_{H^{\pm}}^{2}=\dfrac{1}{4}\left(\lambda_{5}v^{2}+\lambda_{14}\dfrac{v_{\chi}v_{S}}{2s_{2\beta}}\right).
\end{equation}
\subsection{Gauge boson masses $(W_{\mu}^{3},B_{\mu},Z_{\mu}')$}
The $U(1)_{X}$ symmetry generates an additional term in the covariant derivative:
\begin{align}
    D_{\mu}=\partial_{\mu} +igW_{\mu}^{a}T_{a} - ig'\frac{Y}{2}B_{\mu}+ig_{X}Z'_{\mu},
    \label{eq:Covariant_derivative}
\end{align}
and, after symmetry breaking, the $W^{\pm}_{\mu}=(W_{\mu}^{1}\mp  W_{\mu}^{2})/\sqrt{2}$ acquires masses $M_{W}=\frac{gv}{2}$ . The neutral gauge bosons $(W_{\mu}^{3},B_{\mu},Z'_{\mu})$ masses are obtained for the following matrix:
\begin{align}
    M_{0}^{2}=\frac{1}{4}\begin{pmatrix}
    g^{2} v^{2} & -gg'v^{2} & -\frac{2}{3}g g_{X} v^{2}(1+c^{2}_{\beta}) \\
    * & g'{}^{2} v^{2} & \frac{2}{3}g'g_{X} v^{2}(1+c_{\beta}^{2}) \\
    * & * & \frac{4}{9}g_{X}^{2} V_{\chi}^{2}\left[1+(1+3c^{2}_{\beta})\frac{v^{2}}{V_{\chi}^{2}}\right],
    \end{pmatrix}, \nonumber
\end{align}
which have one eigenvalue equal to zero and two eigenvalues corresponding to the masses of the $Z,Z'$ bosons:
\begin{align}
    M_{Z}&\approx\frac{gv}{2\cos{\theta_{W}}},  &    M_{Z'}&\approx \frac{g_{X}v_{\chi}}{3}.
    \label{eq:Gauge_Masses}
\end{align}
 The matrix that diagonalized $M_{0}$ is given in \cite{Mantilla} and has the form:
\begin{equation}
    R_{0}=
    \begin{pmatrix}
    s_{W}&c_{W}&0\\
    c_{W}c_{Z}&-s_{W}c_{Z}&s_{Z}\\
    -c_{W}s_{Z}&s_{W}s_{Z}&c_{Z}
    \end{pmatrix},
\end{equation}
where $\tan\theta_{W}=\frac{g'}{g}$ is the Weinberg angle and $s_{Z}$ is the mixing angle between $Z$ and $Z'$ gauge bosons:
\begin{equation}
    s_{Z}\approx\left(1+s_{\beta}^{2}\right)\dfrac{2g_{X}c_{W}}{3g}\left(\dfrac{m_{Z}}{m_{Z'}}\right)^{2}.
\end{equation}

In order to define the mass eigenstates associated with the Goldstone bosons of the $Z$ and $Z'$ gauge fields (\ref{eq:Gauge_Masses}), it is necessary to use the bilinear terms $Z_{\mu}\partial^{\mu}G_{Z}$ that coming from the kinetic term of the scalar fields. These contributions are expected to be canceled out with the bilinear terms originated in the gauge fixing. The covariant derivative (\ref{eq:Covariant_derivative}) in function of the mass eigenstates can be written as:
\begin{align}
   D_{\mu}=\partial_{\mu}&-\dfrac{ig}{c_{W}}\left(c_{Z}\left(T_{3L}-s_{W}^{2}Q\right)+\dfrac{g_{X}}{g}c_{W}s_{Z}X\right)Z_{\mu}\nonumber\\ 
   &-ig_{X}\left(-\dfrac{g}{g_{X}}\dfrac{s_{Z}}{c_{W
   }}\left(T_{3L}-s_{W}^{2}Q\right)+c_{Z}X\right)Z'_{\mu}.
\end{align}
The gauge-fixing lagrangian in the Feynman gauge is:
\begin{equation}
    \mathcal{L}_{GF}=-\frac{1}{2}\left(\partial_{\mu}Z^{\mu}+M_{Z}G_{Z}\right)^{2}-\frac{1}{2}\left(\partial_{\mu}Z^{'\mu}+M_{Z'}G_{Z'}\right)^{2}.
\end{equation}

Thus, matching the contributions of the covariant derivatives with the bilinear terms from the gauge fixing is possible to obtain the  Goldstone bosons:
\begin{align}
    G_{Z}&=s_{\beta}\eta_{1}+c_{\beta}\eta_{2}+\frac{M_{Z}}{M_{Z'}}s_{Z}\zeta_{\chi},\\
    G_{Z'}&=\zeta_{\chi}-2\frac{v_{1}}{v_{\chi}}\eta_{1}-\frac{v_{2}}{v_{\chi}}\eta_{2}.
\label{eq:eaten-gauge-phases}
\end{align}
The definition of Goldstone bosons allows us to impose new conditions for PQ-charges in order to decouple the axion.
\subsection{PQ coupling to gauge bosons}
Under the $U(1)_{PQ}$ symmetry, the scalar fields transforms as:
\begin{equation}
    \phi_{1}\rightarrow e^{x_{1}\alpha}\phi_{1},\quad 
    \phi_{2}\rightarrow e^{x_{2}\alpha}\phi_{2},\quad
    \chi\rightarrow e^{x_{\chi}\alpha}\chi,\quad
    S\rightarrow e^{x_{S}\alpha}S.
\end{equation}
and the current associated with the PQ-transformation is given by: 
\begin{equation}
    J_{\mu}^{PQ}=x_{S}v_{S}i\partial_{\mu}\zeta_{S}+x_{\chi}v_{\chi}i\partial_{\mu}\zeta_{\chi}+x_{2}v_{2}\partial_{\mu}\eta_{2}+x_{1}v_{1}\partial_{\mu}\eta_{1},
    \label{eq:PQ-current}
\end{equation}
which must be orthogonal with neutral currents at low energy, leading to the following restrictions:
\begin{align}
   0&=v_{1}^{2}x_{1}+v_{2}^{2}x_{2}+(2v_{1}^{2}+v_{2}^{2})\frac{v}{v_{\chi}}x_{\chi},\nonumber\\
   0&=2v_{1}^{2}x_{1}+v_{2}^{2}x_{2}-v_{\chi}^{2}x_{\chi}.
    \label{eq:AxionNorm}
\end{align}
In addition, the $\lambda_{14}$ term in the scalar potential (\ref{eq:scalar-potential}) generates the following equation:
\begin{equation}
x_{\chi}-x_{S}-x_{1}+x_{2}=0.
\label{eq:AxionHiggs}
\end{equation}
So using eqs. (\ref{eq:AxionNorm}), (\ref{eq:AxionHiggs}) and choosing the normalization condition
\begin{equation}
    x_{S}-x_{\chi}=1,
\end{equation}
it is possible to write:
\begin{align}
    x_{1}&=-\dfrac{v_{2}^{2}(v(2v_{1}^{2}+v_{2}^{2})+v_{\chi}^{3})}{v(2v_{1}^{2}+v_{2}^{2})^{2}+v^{2}v_{\chi}^{3}},\qquad x_{\chi}=-\frac{v_{1}^{2}v_{2}^{2}v_{\chi}}{v(2v_{1}^{2}+v_{2}^{2})^{2}+v^{2}v_{\chi}^{3}},\nonumber\\
    x_{2}&=1+x_{1},\qquad\qquad\qquad\qquad\quad x_{S}=1+x_{\chi}.
\end{align}

\section{Fermion sector}
\begin{table}[h]
\centering
\begin{tabular}{cccc|cccc|}
\hline\hline
Quarks	&	$X$	&$PQ$&&	Leptons	&	$X$&$PQ$	\\ \hline 
\multicolumn{7}{c}{SM Fermionic Isospin Doublets}	\\ \hline\hline
$q^{1}_{L}=\left(\begin{array}{c}U^{1} \\ D^{1} \end{array}\right)_{L}$
	&	$+1/3$	&$0$	&&
$\ell^{e}_{L}=\left(\begin{array}{c}\nu^{e} \\ e^{e} \end{array}\right)_{L}$
	&	$0$	&$x_{\ell^{\mu}_{L}}$	\\
$q^{2}_{L}=\left(\begin{array}{c}U^{2} \\ D^{2} \end{array}\right)_{L}$
	&	$0$	&$x_{1}+x_{2}$	&&
$\ell^{\mu}_{L}=\left(\begin{array}{c}\nu^{\mu} \\ e^{\mu} \end{array}\right)_{L}$
	&	$0$	&$-\left(\frac{x_{S}}{2}+x_{2}\right)$		\\
$q^{3}_{L}=\left(\begin{array}{c}U^{3} \\ D^{3} \end{array}\right)_{L}$
	&	$0$	&$0$	&&
$\ell^{\tau}_{L}=\left(\begin{array}{c}\nu^{\tau} \\ e^{\tau} \end{array}\right)_{L}$
	&	$-1$	&$x_2-\frac{x_{S}}{2}+2x_{\chi}$	\\   \hline\hline

\multicolumn{7}{c}{SM Fermionic Isospin Singlets}	\\ \hline\hline
\begin{tabular}{c}$U_{R}^{1,3}$\\$U_{R}^{2}$\\$D_{R}^{1,2}$\\$D_{R}^3$\end{tabular}	&	 
\begin{tabular}{c}$+2/3$\\$+2/3$\\$-1/3$\\$-1/3$\end{tabular}	&
\begin{tabular}{c}$-x_{1}$\\$-x_{2}$\\$2x_{1}+x_{2}$\\$-x_{2}$\end{tabular}	&&
\begin{tabular}{c}$e_{R}^{e}$\\$e_{R}^{\mu}$\\$e_{R}^{\tau}$\end{tabular}	&	
\begin{tabular}{c}$-4/3$\\$-4/3$\\$-1/3$\end{tabular}	&	
\begin{tabular}{c}$-\frac{x_{S}}{2}+2x_{\chi}$\\$-2x_{2}-\frac{x_{S}}{2}$\\$-\frac{x_{S}}{2}+2x_{\chi}$\end{tabular}\\   \hline \hline 

\multicolumn{3}{c}{Non-SM Quarks}	&&	\multicolumn{3}{c}{Non-SM Leptons}	\\ \hline \hline
\begin{tabular}{c}$T_{L}$\\$T_{R}$\end{tabular}	&
\begin{tabular}{c}$+1/3$\\$+2/3$\end{tabular}	&
\begin{tabular}{c}$x_{2}+x_{\chi}$\\$x_{2}$\end{tabular}	&&
\begin{tabular}{c}$\nu_{R}^{e,\mu,\tau}$\\$E_L$\\$E_R$\end{tabular} 	&	
\begin{tabular}{c}$1/3$\\$-1$\\$-2/3$\end{tabular}	&	
\begin{tabular}{c}$-\frac{x_{S}}{2}$\\$-x_{2}+x_{\sigma}$\\$-x_{1}$\end{tabular}\\
$J^{1,2}_{L}$	&	  $0$ 	&$x_{1}-x_{\chi}$	&&	$\mathcal{E}_{L}$	&	$-2/3$	&$-x_{2}+x_{\sigma}$	\\
$J^{1,2}_{R}$	&	 $-1/3$	&$x_{1}$	&&	$\mathcal{E}_{R}$	&	$-1$	&$x_{\sigma}-x_{1}-2x_{2}$	\\ \hline \hline
\end{tabular}
\caption{Non-universal $X$ quantum number and $PQ$ Charge for SM and non-SM fermions.}
\label{tab:Fermionic-content}
\end{table}
In ref. \cite{Mantilla} it was found that considering the $U(1)_{X}$ model with an additional $Z_{2}$ symmetry, it is possible to explain the hierarchical masses of the fermions. In the present work we change the $Z_{2}$-symmetry for a PQ-symmetry that generates the same Yukawa lagrangian and the same ansatz for the masses. Thus, we propose the same Yukawa lagrangian of the ref. \cite{Mantilla} and find a solution for the set of PQ-charges  taking into account the restrictions coming from the scalar potential (\ref{eq:scalar-potential}) and restrictions from axion decoupling at low energies. The most general lagrangian for the quark sector is:

\begin{align}
-\mathcal{L}_Q &=
\overline{q_L^{1}}\left(\widetilde{\phi}_{2}h^{U}_{2}\right)_{12}U_{R}^{2}+\overline{q_L^{1}}\left(\widetilde{\phi}_{2}
h^{T}_{2} \right)_{1}T_{R}+\overline{q_L^{2}}(\widetilde{\phi}_{1} h^{U}_{1})_{22}U_R^{2}
+\overline{q_{L}^{2}} (\widetilde{\phi}_{1} h^{T}_{1})_{2}T_{R}\nonumber\\
&+\overline{q_L^{3}}(\widetilde{\phi}_{1} h^{U}_{1})_{31}U_{R}^{1}+\overline{q_L^{3}}(\widetilde{\phi}_{1}h^{U}_{1})_{33}U_R^{3}+\overline{T_{L}}\left(\chi h_{\chi}^{U}\right)_{j}{U}_{R}^{2}\nonumber\\
&+\overline{T_{L}}\left(\chi h_{\chi}^{T}\right)T_{R}+\overline{q_L^{1}} (\phi_{1}h^{J}_{1})_{11} J^{1}_{R}+\overline{q_L^{1}}(\phi_{1} h^{J}_{1})_{12} J^{2}_{R}\nonumber\\
&+\overline{q_L^{2}}\left(\phi_{2}h^{J}_{2}\right)_{21} J^{1}_{R}
+\overline{q_L^{2}}\left(\phi  _2 h^{J}_{2} \right)_{22}J^{2}_{R}+\overline{q_L^{3}}\left(\phi_{2}h^{D}_{2} \right)_{33}D_R^{3} \nonumber\\
&+\overline{J_{L}^n}\left(\sigma^{*}h_{\sigma }^{D}\right)_{n(1,2)}{D}_{R}^{1,2}+\overline{J_{L}^{n}}\left(\chi ^*h_{\chi }^{J}\right)_{nm}{J}_{R}^{m}+h.c.,
 \label{eq:Lagrangian-quark-sector}
\end{align}
and the lagrangian for  the neutral and charged leptonic sector has the following structure:
\begin{align}
-\mathcal{L}_{Y,E} &= 
\left(g_{e\mu}^{2e}\overline{\ell^{e}_{L}}\phi_{2}e^{\mu}_{R} 
+g_{\mu\mu}^{2e}\overline{\ell^{\mu}_{L}}\phi_{2}e^{\mu}_{R} +g_{\tau e}^{2e}\overline{\ell^{\tau}_{L}}\phi_{2}e^{e}_{R} +g_{\tau\tau}^{2e}\overline{\ell^{\tau}_{L}}\phi_{2}e^{\tau}_{R} 
+g_{Ee}^{1}\overline{\ell^{e}_{L}}\phi_{1}E_{R}\right. \nonumber\\
&+g_{E\mu}^{1}\overline{\ell^{\mu}_{L}}\phi_{1}{E}_{R}
+h^{\sigma e}_{E}\overline{E_{L}}\sigma^{*} e^{e}_{R} 
+ h^{\sigma\mu}_{\mathcal{E}}\overline{\mathcal{E}_{L}}\sigma e^{\mu}_{R} 
+ h^{\sigma\tau}_{E}\overline{E_{L}}\sigma^{*} e^{\tau}_{R} + h^{\chi E}\overline{E_{L}}\chi E_{R}\nonumber\\
&\left.+h^{\chi\mathcal{E}}\overline{\mathcal{E}_{L}}\chi^{*} \mathcal{E}_{R} + \mathrm{h.c.}\right)
+h_{2e}^{\nu e}\overline{\ell^{e}_{L}}\tilde{\phi}_{2}\nu^{e}_{R} 
+ h_{2e}^{\nu \mu}\overline{\ell^{e}_{L}}\tilde{\phi}_{2}\nu^{\mu}_{R} 
+ h_{2e}^{\nu \tau}\overline{\ell^{e}_{L}}\tilde{\phi}_{2}\nu^{\tau}_{R} 
+ h_{2\mu}^{\nu e}\overline{\ell^{\mu}_{L}}\tilde{\phi}_{2}\nu^{e}_{R} \nonumber\\
&+h_{2\mu}^{\nu \mu}\overline{\ell^{\mu}_{L}}\tilde{\phi}_{2}\nu^{\mu}_{R} 
+ h_{2\mu}^{\nu \tau}\overline{\ell^{\mu}_{L}}\tilde{\phi}_{2}\nu^{\tau}_{R}
+h_{S i}^{\nu e} \overline{\nu_{R}^{i\;C}} S\nu_{R}^{e}
+h_{S i}^{\nu\mu} \overline{\nu_{R}^{i\;C}}S\nu_{R}^{\mu}
+h_{S i}^{\nu\tau} \overline{\nu_{R}^{i\;C}} S\nu_{R}^{\tau}.
\label{eq:Leptonic-Lagrangian}
\end{align}
The restrictions on the set of PQ charges in order to obtain the lagrangians with the appropriate ansatz to be able to generate masses in the fermionic sector is given in table (\ref{tab:Fermionic-content}). In addition, the restrictions imposed by eqs. (\ref{eq:scalar-potential}), (\ref{eq:Lagrangian-quark-sector}) and (\ref{eq:Leptonic-Lagrangian}), implies a PQ charge for the Higgs singlet $\sigma$ is:
\begin{equation}
     x_\sigma=x_{1}+x_{2}+x_{\chi}.
\end{equation}
Then, it is possible to build the rotated mass squared matrices for the up and down sector through a see saw mechanism:

\begin{align}
\mathbbm{m}_U^{2}&=\left(V_{L}^{(U)}\right)^{T} \mathbb{M}_{U}^{2} V_{L}^{(U)}=
\begin{pmatrix}
m_U^2 & 0 \\
0 & m_T^2
\end{pmatrix},\quad
\mathbbm{m} _D^2=\left(V_{L}^{(D)}\right)^T \mathbb{M}_{D}^{2} V_{L}^{(D)}=
\begin{pmatrix}
m_U^2 & 0 \\
0 & m_J^2
\end{pmatrix},\\
m_U^2 &\approx  \frac{1}{2}
\begin{pmatrix}
v_2^2  r_1^2 & v_1 v _2 r_1r_2 & 0 \\
v _1 v _2  r_1r_2 & v _1^2  r_2^2 & 0 \\
0 & 0 & v _1 ^2 \left(a_{31}^2+a_{33}^2 \right)
\end{pmatrix},
\quad
m_D^2=\frac{1}{2}\begin{pmatrix}
0 & 0 &  0    \\
0 & 0 & 0  \\
0 & 0 & B_{33}^{2}v_{2}^{2} 
\end{pmatrix},\\
m_{T}^{2}&=\frac{v_{\chi}^{2}}{2}\left((c_{2}^{\chi})^{2}+(d^{T\chi})^{2}\right),
\hspace{3.9cm}
m_{J}^{2}=\frac{v_{\chi}^{2}}{2}
\begin{pmatrix}
k_{11}^{2}&0\\
0&k_{22}^{2}
\end{pmatrix}.
\end{align}

\begin{figure}[h]
\centering
\includegraphics[scale=0.6]{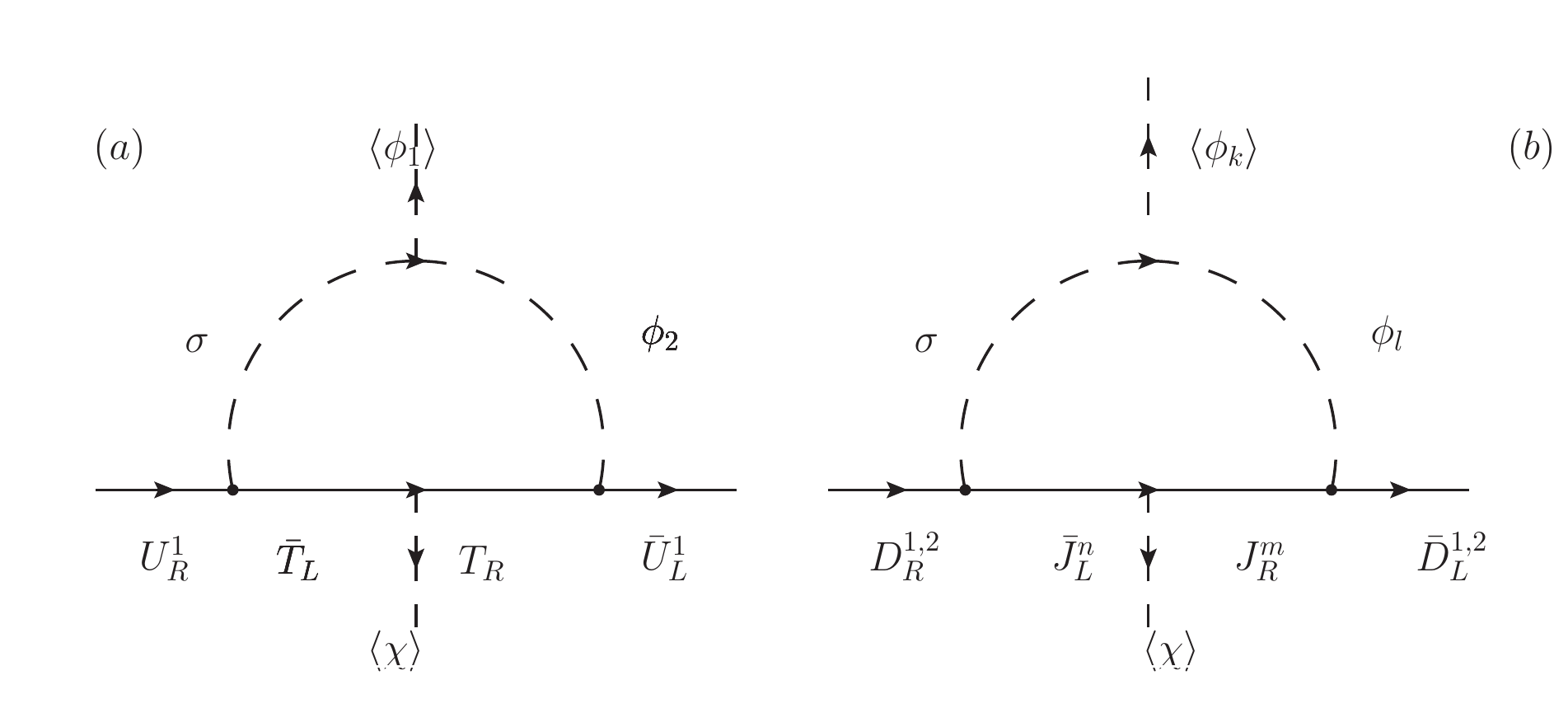}
\caption{Mass $1$-loop corrections  for the (a) up, (b)down sector}
\label{fig:Radiative Corrections}
\end{figure}

The $u,d$ and $s$ quarks turn out to be massless, so it is necessary to introduce radiative corrections to the model. These corrections are shown in the Figure (\ref{fig:Radiative Corrections}) in the diagrams $(a)$ for the up sector, which adds small contributions to the $m_U^2$ matrix through the term:
\begin{small}
\begin{equation}
\Sigma _{11}=\frac{-1}{16\pi ^2}\frac{f'\left(h^U_{\sigma}\right)_1\left(h^T_2\right)_1}{\sqrt{2}M_T}C_0\left(\frac{M_2}{M_T},\frac{M_{\sigma}}{M_T}\right),
\end{equation} 
\end{small}
where:
\begin{small}
\begin{equation}
C_0\left(x_1,x_2\right)=\frac{1}{\left(1-x_1^2\right)\left(1-x_2^2\right)\left(x_1^2-x_2^2\right)}\left[x_1^2x_2^2\ln\left(\frac{{x_1^2}}{x_2^2}\right)-x_1^2\ln x_1^2+x_2^2 \ln x_2^2\right],
\label{oneloop-coef}
\end{equation}
\end{small}
and $(b)$ for down sector, through the coupling with the scalar singlet $\sigma$ generating a one-loop correction of the form:

\begin{equation}
\Sigma _{lj}=\frac{-1}{16\pi ^2}\frac{f'\left(h^J_l\right)_{lm}\left(h^D_{\sigma }\right)_{nj}}{\sqrt{2}M_J}C_0\left(\frac{M_l}{M_J},\frac{M_{\sigma}}{M_J}\right).
\end{equation}

In relation to the charged leptonic sector, the hierarchy obtained in the squared mass matrix can be analyzed again through the see saw mechanism. The PQ charges configuration allow us to build the extended mass matrix for the charged leptons in the following way:

\begin{equation}
   \mathcal{M}_{LC}=
   \left( \begin{array}{cccccc}
        \Sigma_{11}&g^{2e}_{e\mu}&\Sigma_{13}&|&g^{1}_{Ee}&0\\
        \Sigma_{21}&g^{2e}_{\mu\mu}&\Sigma_{23}&|&g^{1}_{E\mu}&0\\
        g^{2e}_{\tau e}&0&g^{2e}_{\tau\tau}&|&0&0\\
        -&-&-&-&-&-\\
        0&0&0&|&h^{\chi E}&0\\
        0&0&0&|&0&h^{\chi\mathcal{E}}
    \end{array}
    \right).
\end{equation}

\begin{figure}
\centering
\includegraphics[scale=0.6]{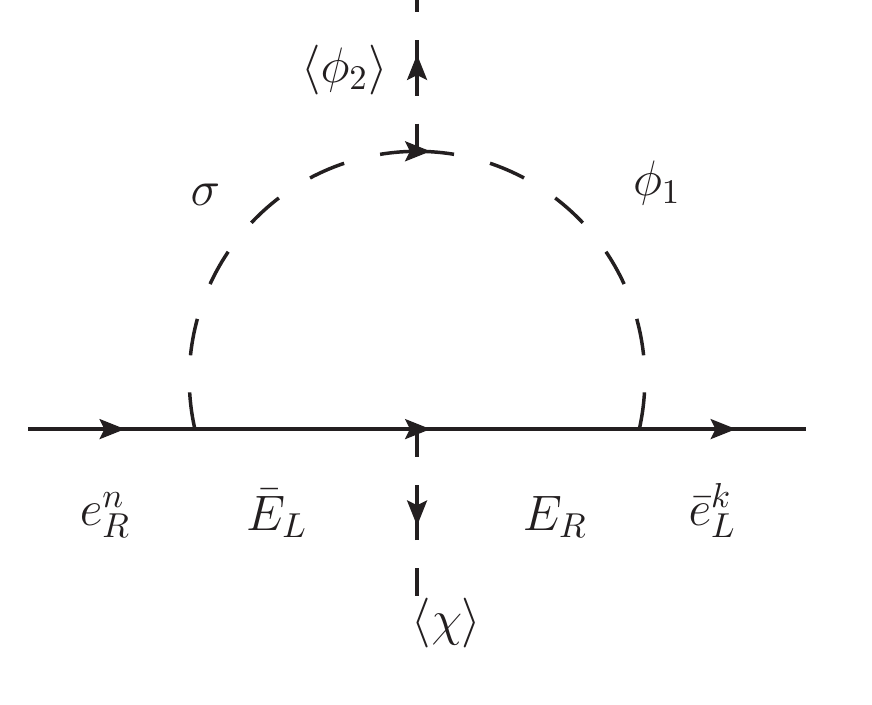}
\caption{Mass $1$-loop corrections for the electron}
\label{fig:RC-electron}
\end{figure}

In the same way as in the model \cite{Mantilla}, the electron does not acquire mass at tree level, so is necessary to implement radiative correction in order to produce finite mass, that corresponds to the $\Sigma_{ij}$ entries in the $ \mathcal{M}_{LC}$ matrix associated to the Figure (\ref{fig:RC-electron}), taking in account the interactions with the $\sigma$-singlet.

For the neutral sector, assuming the condition of the existence of the right-handed neutrino fields, the mass term in the basis $N=\left(\nu^{e,\mu,\tau}_{L},\left(\nu_{R}^{e,\mu,\tau}\right)^{C}\right)^{T}$ has the form:
\begin{equation}
    -\mathcal{L}_{N}=\frac{1}{2}\bar{N^{C}}M_{\nu}N_{L};\quad    M_{\nu}=
    \begin{pmatrix}
        0&m_{D}^{T}\\
        m_{D}&M_{S}
    \end{pmatrix}.
\end{equation}
From eq. (\ref{eq:Leptonic-Lagrangian}) it is possible to see that after SSB, the right-handed neutrinos get masses \cite{3}, proportional to:
\begin{equation}
   h^{\nu i}_{Sj} \frac{v_{S}}{\sqrt{2}},
\end{equation}
where $i,j=e,\mu,\tau$ and the Dirac masses are given by:
\begin{equation}
    m_{D}=h_{2a}^{\nu i}\frac{v_{2}}{\sqrt{2}},
\end{equation}
with $i=e,\mu,\tau$ and $a=e,\mu$. The last row associated with $\tau$ is forbidden by the PQ-symmetry, generating a massless active neutrino. Then, under the see-saw mechanism, the light masses are of the form $m_{\nu}\equiv M_{D}^{T}M_{S}M_{D}\sim\mathcal{O}\left(\frac{v_{2}^{2}}{v_{S}}\right)$. For instance, if $v_{2}\sim m_{\tau}$ and $v_{S}\sim 10^{10}GeV$, the lighter neutrinos has a mass in the order of $\sim\mathcal{O}(eV)$.

The diagonalization of $m_{\nu}$ determines the mass eigenvalues where the lightest state has mass equals to zero, and the other two are associated with squared mass differences $\Delta m_{12}^{2},\Delta m_{23}^{2}$ depend on the Yukawa couplings  \cite{8}. The mixing $\theta$-angles that produces the PMNS matrix are obtained in the same way as in \cite{Mantilla}, where we identified $\frac{\mu_{N}}{v_{\chi}^{2}}$, coming from the inverse see-saw mechanism, with $\frac{1}{v_{S}}$. Using experimental data is possible to find the allowed parameter regions for the Yukawa couplings $h_{2a}^{\nu i}$ consistent with the neutrino oscillations \cite{Maki} which are shown in tables IV and V in \cite{Mantilla}.

\section{Conclusions}
The introduction of the $U(1)_{X}$ and the additional $U(1)_{PQ}$ symmetry to the SM allow to understand three fundamental problems. The model provides an elegant solution to the strong CP-problem,  also generates a correct hierarchy for the masses of the fermionic sector and gives a new scale to the see-saw mechanism, generating masses for the active neutrinos. 


\end{document}

%% file: econfmacros.tex
\def\beq{\begin{equation}}
\def\eeq#1{\label{#1}\end{equation}}
\def\eeqn{\end{equation}}

\def\beqa{\begin{eqnarray}}
\def\eeqa#1{\label{#1}\end{eqnarray}}
\def\eeqan{\end{eqnarray}}

\let\bar=\overbar

\def\Dslash{\not{\hbox{\kern-4pt $D$}}}
\def\dslash{\not{\hbox{\kern-2pt $\del$}}}

\def\msb{{\bar{\ssstyle M \kern -1pt S}}}

